\documentclass[showpacs]{revtex4}

\usepackage{amssymb}
\usepackage{amsmath}
\usepackage{graphicx}
\usepackage{multirow}
\usepackage{caption}
\usepackage{subfig}
\usepackage{float}
\begin{document}
\title
{Magnetohydrodynamic stability of stochastically driven accretion flows}
\author
{Sujit Kumar Nath$^1$, Banibrata Mukhopadhyay$^1$,
Amit K. Chattopadhyay$^{2}$}
\address{1. 
Department of Physics, Indian Institute of Science, Bangalore 560 012, India;\\
sujitkumar@physics.iisc.ernet.in ;\,\,bm@physics.iisc.ernet.in \\
2. Aston University, Non-linearity and Complexity Research Group, Engineering \\
and Applied Science, Birmingham B4 7ET, UK; a.k.chattopadhyay@aston.ac.uk
}


\begin{abstract}

We investigate the evolution of magnetohydrodynamic/hydromagnetic perturbations in the presence of stochastic noise
in rotating shear flows. The particular emphasis
is the flows whose angular velocity decreases but specific angular momentum increases with increasing
radial coordinate. Such flows, however, are Rayleigh stable, but must be turbulent in 
order to explain astrophysical observed data and, hence, reveal a mismatch between the linear theory and 
observations/experiments. The mismatch seems to have been resolved, atleast in certain regimes,
in the presence of weak magnetic field revealing magnetorotational instability. 
The present work explores the effects of stochastic noise on such magnetohydrodynamic flows, in order 
to resolve the above mismatch generically for the hot flows. 
We essentially concentrate on a small section of such a flow which is nothing but a plane shear 
flow supplemented by the Coriolis effect, mimicking a small section of an astrophysical accretion disk
around a compact object.
It is found that such stochastically driven flows exhibit large temporal and 
spatial auto-correlations and cross-correlations of perturbation and hence large energy 
dissipations of perturbation, which generate instability. 
Interestingly, auto-correlations and cross-correlations appear independent of 
background angular velocity profiles, which are Rayleigh stable, indicating their {\it universality}.
This work, to the best of our knowledge, is the first attempt to understand the evolution of
three-dimensional hydromagnetic perturbations in rotating shear flows in the presence of 
stochastic noise.

\end{abstract}


\maketitle

\vspace{0.2cm}

\section{Introduction}


Recently, Mukhopadhyay \& Chattopadhyay \cite{mc13} (see, the references therein) have initiated exploring effects of
stochastic noise in rotating shear flows in three dimensions with particular emphasize to astrophysical
accretion disks. They have essentially addressed the evolutions of pure hydrodynamic perturbations
and found them to be adequate enough to explain instability and subsequent turbulence
therein. This is in accordance with the fact that in three dimensions, one requires to invoke 
extra physics to reveal large energy growth or even instability in the system \cite{bmraha}.
This is very important for charge natural flows like accretion disks around quiescent cataclysmic variables,
in protoplanetary and star-forming disks,
and the outer region of disks in active galactic nuclei etc. where flows are cold and of
low ionization and effectively neutral in charge. 

In the cases of hot flows, e.g. disks around black holes, magnetorotational instability is generally 
believed to be responsible for turbulence and hence transport of angular momentum therein.
The problem has been well studied and has had a long history of fluid mechanical insight into the 
rotating shear flows and subsequently accretion disk problem in the linearly stable regime, when origin of 
turbulence is a major issue \cite{gu,kim,mk,man,rud,dau,zahn,klar,dub1,dub2}.
Based on `shearing sheet' approximation, without \cite{balbusetal96,hawleyetal99}
and with \cite{ll05} explicit viscosity, some authors attempted to tackle the issue of turbulence
in hot accretion disks. However, other authors argued for 
limitations in this work \cite{pumir96,fromang-papaloizou07}. While the authors, who did 
not include explicit viscosity,
could not directly define a Reynolds number ($R_e$),
their estimated $R_e$ from the simulations is $\sim 10^3 -10^4$.
They also did not find any evidence for a subcritical
transition to turbulence. Based on the simulations including explicit viscosity,
other authors could achieve $R_e \approx 4 \times 10^4$, and concluded that Keplerian like 
flows could exhibit very weak turbulence, particularly in absence of magnetic field. 
However, the recent experimental 
results by Paoletti et al. \cite{paoletti}, clearly argue for the significant level of transport 
from hydrodynamics alone. Moreover, the results from direct numerical simulations 
\cite{avila} also argue for hydrodynamic instability and turbulence at low $R_e$.

In the present paper, we extend the work by Mukhopadhyay \& Chattopadhyay \cite{mc13} and
investigate the amplification of linear magnetohydrodynamic/hydromagnetic 
perturbations in Rayleigh stable rotating, hot, shear flows in the presence of stochastic noise in three dimensions, 
leading to instability and plausible turbulence. The earlier paper \cite{mc13} already summarized
the association of growing, unstable modes generated by perturbed 
flows with statistical physics, in particular effects of noise in such flows, 
based on which the present work has been founded. 
Hence we do not repeat them here. 
The effects of white noise in a linear, nonrotating, non-magnetized shear flow, which is non-normal in nature, 
was also studied by earlier authors \cite{ep03}. 
In the present study, we implement the ideas of statistical physics, already implemented by above authors,
to rotating, magnetized, shear flows 
in order to obtain the correlation
energy growths of fluctuation/perturbation and underlying scaling properties. 

In the next section, we first recall the equations describing the stochastically forced perturbed
flows, namely magnetized version of the set of Orr-Sommerfeld and Squire equations proposed
by Mukhopadhyay \& Chattopadhyay \cite{mc13} in the presence of noise, which are to be solved for the present purpose. 
Subsequently, in \S 3 we investigate
the temporal and spatial auto-correlations and cross-correlations of perturbation in the presence 
of white noise in detail, in order to understand the
plausible instability in the flows. In \S 4, we study the correlations in the presence of colored noise.
Finally, we summarize the results with conclusions in \S 5.

\section{Equations describing perturbed magnetized rotating shear flows in the presence of noise}

The linearized Navier-Stokes equation in the presence of background plane shear $(0,-x,0)$ and magnetic field $(0,B_1,1)$,
when $B_1$ being a constant, both expressed in the dimensionless units, 
in the presence of background angular velocity profile
$\Omega\propto r^{-q}$, when $r$ being the distance from the center of the system,
in a small section of the incompressible flow with $-l/2\le x\le l/2$, has already been established \cite{mc13}. 
The underlying equations
are nothing but the linearized set of hydromagnetic equations including the equations of induction 
in a local Cartesian coordinate. Here, we plan to work with the dimensionless variables, when any
length is expressed in units of the size of system $L$ in the $x-$direction, the time in
units of the inverse of background angular velocity of the flow about $z-$direction $\Omega$,
the velocity in $q\Omega L$ ($1\le q<2$), and other variables are expressed accordingly 
(see, e.g., \cite{man,bmraha,mc13} for detailed description of the
choice of coordinate in a small section). Hence, in dimensionless units, the set of equations is given by
\begin{eqnarray}
\left(\frac{\partial }{\partial t}-x\frac{\partial }{\partial y}\right)u-\frac{2v}{q}+\frac{\partial p_{\rm tot}}
{\partial x}-\frac{1}{4\pi}\left(B_1\frac{\partial B_x}{\partial y}+\frac{\partial B_x}{\partial z}\right)
=\frac{1}{R_e}\nabla^2u,
\label{u}
\end{eqnarray}
\begin{eqnarray}
\left(\frac{\partial }{\partial t}-x\frac{\partial }{\partial y}\right)v+\left(\frac{2}{q}-1\right)u+\frac{\partial p_{\rm tot}}
{\partial y}-\frac{1}{4\pi}\left(B_1\frac{\partial B_y}{\partial y}+\frac{\partial B_y}{\partial z}\right)
=\frac{1}{R_e}\nabla^2v,
\label{v}
\end{eqnarray}
\begin{eqnarray}
\left(\frac{\partial }{\partial t}-x\frac{\partial }{\partial y}\right)w+\frac{\partial p_{\rm tot}}
{\partial z}-\frac{1}{4\pi}\left(B_1\frac{\partial B_z}{\partial y}+\frac{\partial B_z}{\partial z}\right)
=\frac{1}{R_e}\nabla^2w,
\label{w}
\end{eqnarray}
\begin{eqnarray}
\frac{\partial B_x}{\partial t}=\frac{\partial u }{\partial z}+ B_1\frac{\partial u}{\partial y}+x\frac{\partial B_x}
{\partial y}+\frac{1}{R_m}\nabla^2B_x,
\label{bx}
\end{eqnarray}
\begin{eqnarray}
\frac{\partial B_y}{\partial t}=\frac{\partial v }{\partial z}+ B_1\frac{\partial v}{\partial y}-x\frac{\partial B_x}
{\partial x}-x\frac{\partial B_z}{\partial z}-B_x+\frac{1}{R_m}\nabla^2B_y,
\label{by}
\end{eqnarray}
\begin{eqnarray}
\frac{\partial B_z}{\partial t}=\frac{\partial w }{\partial z}+ B_1\frac{\partial w}{\partial y}+x\frac{\partial B_z}
{\partial y}+\frac{1}{R_m}\nabla^2B_z,
\label{bz}
\end{eqnarray}
when the vectors for velocity and magnetic field perturbations are $(u,v,w)$ and $(B_x,B_y,B_z)$ respectively,
$R_e$ and $R_m$ are the hydrodynamic and magnetic Reynolds numbers respectively, $p_{\rm tot}$ is the total
pressure perturbation (including that due to the magnetic field). Above equations are
supplemented by the conditions for incompressibility and absence of magnetic charge, given respectively by
\begin{eqnarray}
\frac{\partial u}{\partial x}+\frac{\partial v }{\partial y}+ \frac{\partial w}{\partial z}=0,
\label{eoc}
\end{eqnarray}
\begin{eqnarray}
\frac{\partial B_x}{\partial x}+\frac{\partial B_y }{\partial y}+ \frac{\partial B_z}{\partial z}=0.
\label{monop}
\end{eqnarray}
Now the above equations in the presence of stochastic noise 
can be recasted into magnetized version of Orr-Sommerfeld and Squire equations in the presence of the Coriolis force, 
given by \cite{mc13}
\begin{equation}
\left(\frac{\partial}{\partial t}-x\frac{\partial}{\partial y}\right)\nabla^2 u
+\frac{2}{q}\frac{\partial \zeta}{\partial z}-\frac{1}{4\pi}\left(B_1\frac{\partial}{\partial y}+\frac{\partial}{\partial z}\right)\nabla^2B_x
=\frac{1}{R_e}\nabla^4 u+\eta_1(x,t),
\label{orrv}
\end{equation}
\begin{equation}
\left(\frac{\partial}{\partial t}-x\frac{\partial}{\partial y}\right)\zeta
+\frac{\partial u}{\partial z}
-\frac{2}{q}\frac{\partial u}{\partial z}-\frac{1}{4\pi}\left(B_1\frac{\partial}{\partial y}+\frac{\partial}{\partial z}\right)\zeta_B=\frac{1}{R_e}\nabla^2 \zeta +
\eta_2(x,t),
\label{zeta}
\end{equation}
\begin{eqnarray}
\left(\frac{\partial }{\partial t}-x\frac{\partial}{\partial y}\right)B_x
-B_1\frac{\partial u}{\partial y}-\frac{\partial u}{\partial z}=
\frac{1}{R_m}\nabla^2B_x+\eta_3(x,t),
\label{orrb}
\end{eqnarray}
\begin{eqnarray}
\left(\frac{\partial }{\partial t}-x\frac{\partial}{\partial y}\right)\zeta_B-
\frac{\partial \zeta}{\partial z}-B_1\frac{\partial \zeta}{\partial y}-
\frac{\partial B_x}{\partial z}=\frac{1}{R_m}\nabla^2\zeta_B+\eta_4(x,t).
\label{orrbzeta}
\end{eqnarray}
where $\eta_{1,2,3,4}$ are the components of noise arising in the linearized system due
to stochastic perturbation such that $<\eta_i(\vec x,t) \eta_j(\vec x',t')>=D_i(\vec x)\:\delta^3(\vec x-\vec x')\:\delta(t-t')\:\delta_{ij}$.
The long time, large distance behaviors of the correlations of noise are encapsulated in 
$D_i(\vec x)$ which is a structure pioneered by Forster, Nelson \& Stephen \cite{nelson}. 
In the Fourier space, however,
the structure of the correlation function $D_i(\vec k)$
depends on the regime under consideration.
It can be shown for all (non-linear) non-inertial flows 
\cite{nelson,akc_shear} that $D_i(k) \sim 1/k^d$, where $d$ is the spatial dimension, without
vertex correction and $D_i(k) \sim 1/k^{d-\alpha}$, with $\alpha>0$, in the presence of vertex correction.
Note, however, that $D_i(\vec x)$ is constant for white noise.

As before \cite{mc13} we focus onto the narrow gap limit, 
where in a local analysis we consider a small radially confined region
of the flow, while the azimuthal and vertical confinements are imposed by a 
periodic boundary conditions accordingly
in such a way that the perturbation wave-vector can be assumed to be isotropic.
Note that it could be easily 
extended to a free-slip case. The only modification this would bring about  
is in the values of the limits (now finite, 
instead of infinite). Apart from complicating the 
calculation of the resultant integrals which will have poles of different 
nature in different ranges of $k$, this would not serve in bringing
any practical change for the present purpose. For further details, see Mukhopadhyay \& Chattopadhyay \cite{mc13}.
Hence, we can resort to a Fourier series expansion of
$u$, $\zeta$, $B_x$, $\zeta_B$ and $\eta_i$ as
\begin{eqnarray}
\nonumber
u(\vec{x},t)=\int\tilde{u}_{\vec{k},\omega}\,e^{i(\vec{k}.\vec{x}-\omega t)}d^3k\,d\omega,\\
\nonumber
\zeta(\vec{x},t)=\int\tilde{\zeta}_{\vec{k},\omega}\,e^{i(\vec{k}.\vec{x}-\omega t)}d^3k\,d\omega, \\
\nonumber
B_x(\vec{x},t)=\int\tilde{B_x}_{\vec{k},\omega}\,e^{i(\vec{k}.\vec{x}-\omega t)}d^3k\,d\omega,\\
\nonumber
\zeta_B(\vec{x},t)=\int\tilde{\zeta_B}_{\vec{k},\omega}\,e^{i(\vec{k}.\vec{x}-\omega t)}d^3k\,d\omega,\\
\eta_i(\vec x,t) = \int\tilde{\eta_i}_{\vec{k},\omega}\,e^{i(\vec{k}.\vec{x}-\omega t)}d^3k\,d\omega,
\label{four}
\end{eqnarray}
and substituting them into equations (\ref{orrv}), (\ref{zeta}), (\ref{orrb}) and (\ref{orrbzeta}) we obtain 
\begin{eqnarray}
\left(\begin{array}{cr}\tilde{u}_{\vec{k},\omega}\\ 
\tilde{\zeta}_{\vec{k},\omega}\\\tilde{B}_{x_{\vec{k},\omega}}\\\tilde{\zeta_B}_{\vec{k},\omega}\end{array}\right)={\cal M}^{-1}\left(\begin{array}{cr}\tilde{\eta_1}_{\vec{k},\omega}\\ 
\tilde{\eta_2}_{\vec{k},\omega}\\\tilde{\eta_3}_{\vec{k},\omega}\\\tilde{\eta_4}_{\vec{k},\omega}\end{array}\right),
\label{mat1}
\end{eqnarray}
where 
\begin{eqnarray}
{\cal M}=\left(\begin{array}{cr}{\cal M}_{11}\,\,\,\,\, {\cal M}_{12}\,\,\,\,\,{\cal M}_{13}\,\,\,\,\,{\cal M}_{14}\\ 
{\cal M}_{21}\,\,\,\,\, {\cal M}_{22}\,\,\,\,\,{\cal M}_{23}\,\,\,\,\,{\cal M}_{24}\\
{\cal M}_{31}\,\,\,\,\, {\cal M}_{32}\,\,\,\,\,{\cal M}_{33}\,\,\,\,\,{\cal M}_{34}\\
{\cal M}_{41}\,\,\,\,\, {\cal M}_{42}\,\,\,\,\,{\cal M}_{43}\,\,\,\,\,{\cal M}_{44}\end{array}\right),
\label{mat2}
\end{eqnarray}
\begin{eqnarray}
\nonumber
&&{\cal M}_{11}=ik^2\omega+ilk^2k_y-\frac{k^4}{R_e},\hspace{2mm}{\cal M}_{12}=\frac{2ik_z}{q},\hspace{2mm}{\cal M}_{13}=\frac{ik^2}{4\pi}(B_1k_y+k_z),\hspace{2mm}{\cal M}_{14}=0,\\
\nonumber
&&{\cal M}_{21}=ik_z\left(1-\frac{2}{q}\right),\hspace{2mm}{\cal M}_{22}=-i\omega-ilk_y+\frac{k^2}{R_e},\hspace{2mm}{\cal M}_{23}=0,\hspace{2mm}{\cal M}_{24}=\frac{-i}{4\pi}\left(B_1k_y+k_z\right),\\
\nonumber
&&{\cal M}_{31}=\left(-iB_1k_y-ik_z\right),\hspace{2mm}{\cal M}_{32}=0,\hspace{2mm}{\cal M}_{33}=\left(-i\omega-ilk_y+\frac{k^2}{R_m}\right),\hspace{2mm}{\cal M}_{34}=0,\\
&&{\cal M}_{41}=0,\hspace{2mm}{\cal M}_{42}=\left(-iB_1k_y-ik_z\right),\hspace{2mm}{\cal M}_{43}=-ik_z,\hspace{2mm}{\cal M}_{44}=\left(-i\omega-ilk_y+\frac{k^2}{R_m}\right),
\label{matrixcoeff}
\end{eqnarray}
when $\tilde{{\eta}_{i}}_{\vec{k},\omega}$; $i=1,2,3,4$, are the components of noise in $k-\omega$ space,
$k=\sqrt{k_x^2+k_y^2+k_z^2}$. See \cite{mc13} for other details.

\section{Two-point correlations of perturbation in the presence of white noise}

We now look at the spatio-temporal auto-correlations and cross-correlations 
of the perturbation flow fields $u$, $\zeta$, $B_x$ and $\zeta_B$ for very large $R_e$ and 
$R_m$ \cite{barabasi_stanley}. This choice of large $R_e/R_M$ is quite meaningful for 
astrophysical flows. 
For the present purpose, the magnitudes and gradients (scalings) of these correlations of perturbations would 
plausibly indicate noise induced instability which could lead to 
turbulence in rotating shear flows.

\subsection{Temporal correlation}
\subsubsection{Auto-correlations}

Assuming $<\tilde{\eta_i}_{\vec{k},\omega}\,\tilde{\eta_j}_{-\vec{k},-\omega}>=\delta_{ij}$, without loss of any important physics,
we obtain the temporal correlations of velocity,
vorticity, magnetic field and magnetic vorticity perturbations given below as
\begin{eqnarray}
\nonumber
&&<u(\vec{x},t)\,u(\vec{x},t+\tau)>=C_u(\tau)=\int d^3k\,d\omega\,e^{-i\omega\tau}<\tilde{u}_{\vec{k},\omega}\,
\tilde{u}_{-\vec{k},-\omega}>\\
\nonumber
&&<\zeta(\vec{x},t)\,\zeta(\vec{x},t+\tau)>=C_\zeta(\tau)=\int d^3k\,d\omega\,e^{-i\omega\tau}<\tilde{\zeta}_{\vec{k},\omega}\,
\tilde{\zeta}_{-\vec{k},-\omega}>\\
\nonumber
&&{\hskip-1.3cm <B_x(\vec{x},t)\,\ B_x(\vec{x},t+\tau)}>
=C_{B_x}(\tau)
=\int d^3k\,d\omega\,e^{-i\omega\tau}<\tilde{B_x}_{\vec{k},\omega}\,
\tilde{B_x}_{-\vec{k},-\omega}>\\
&&{\hskip-1.3cm <\zeta_B(\vec{x},t)\,\,\zeta_B(\vec{x},t+\tau)}>
=C_{\zeta_B}(\tau)
=\int d^3k\,d\omega\,e^{-i\omega\tau}<\tilde{\zeta_B}_{\vec{k},\omega}\,
\tilde{\zeta_B}_{-\vec{k},-\omega}>.
\label{tempautocorr}
\end{eqnarray}
We further consider the projected hyper-surface for which $k_x=k_y=k_z=k/\sqrt{3}$, without much loss of
generality for the present purpose. This corresponds to a special choice of initial perturbation.
As our one of the major interests is to understand
the scaling laws, this restriction would not matter, which however may affect the magnitude of 
the correlations. This further helps in introducing the incompressibility
constraints on the noise in the corresponding representation easily, which 
becomes independent of $k$ (see \cite{ep03} for details).

We now perform the $\omega$-integration of the integrands in equation (\ref{tempautocorr}) by computing 
the four second order poles of the kernel which are functions of $k$. 
The form of all the integrands in equation (\ref{tempautocorr}) is given by
\begin{eqnarray}
\nonumber
f(k,\omega)=\frac{p(k,\omega)}{[\omega-\omega_1(k)]^2[\omega-\omega_2(k)]^2[\omega-\omega_3(k)]^2[\omega-\omega_4(k)]^2},
\end{eqnarray}
which clearly reveals second order poles at $\omega_1, \omega_2, \omega_3$ and $\omega_4$. 
We choose the range of $k$ in such a way that the poles lie in the upper-half of the complex plane.
Then by summing up the residues at the appropriate poles, we evaluate the magnitude of 
frequency part of the integration. 
Finally, integrating the rest from $k_0$ to $k_m$, where $k_0=2\pi/l_{\rm max}$,
$k_m=2\pi/l_{\rm min}$ and $l=l_{\rm max}-l_{\rm min}$, 
being the size of the chosen small section of the flow in the radial direction 
(chosen to be $2$ throughout for the present calculations),
we obtain $C_u(\tau)$, $C_\zeta(\tau)$, $C_{B_x}(\tau)$ and $C_{\zeta_B}(\tau)$.

\begin{figure}[H]
 \centering
\includegraphics[scale=0.9]{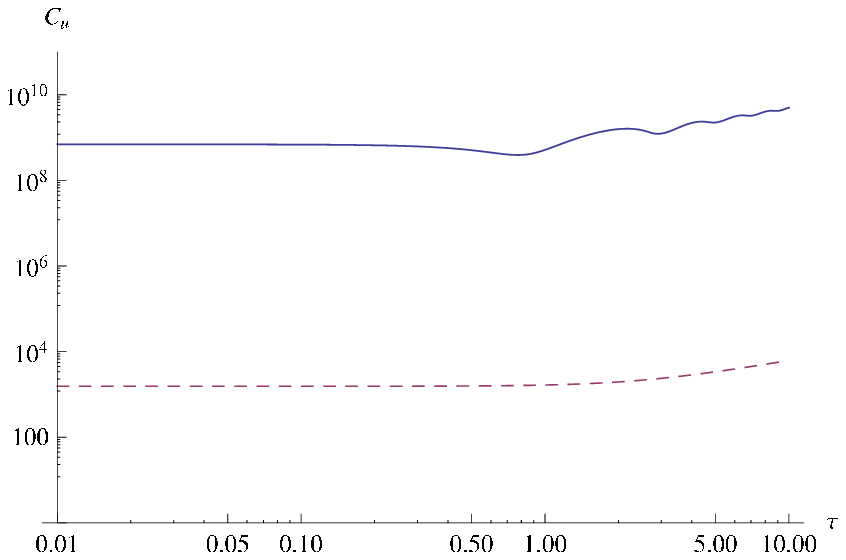}
\caption{Comparing temporal auto-correlations of velocity, when $q=1.5$, between magnetic (solid line) and 
non-magnetic (dashed line) flows.
}
\label{tempcomq1p5}
\end{figure}

\begin{figure}[H]
 \centering
\includegraphics[scale=0.9]{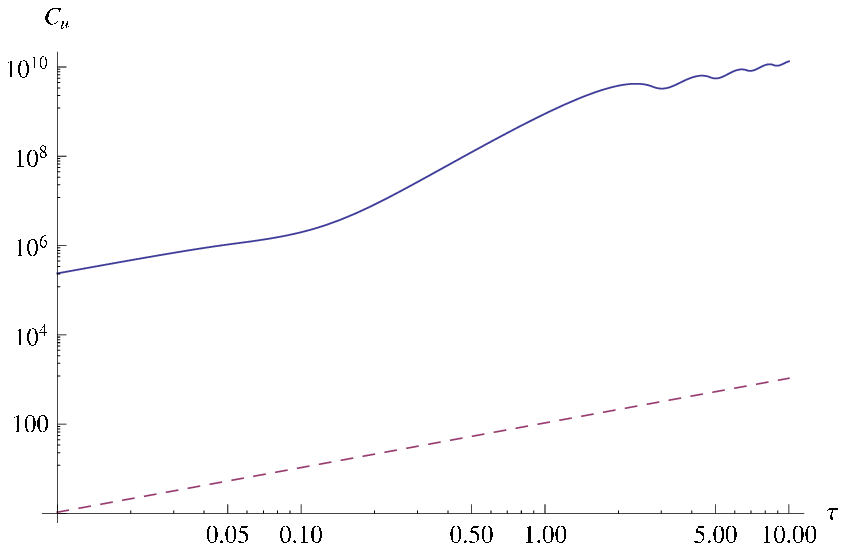}
\caption{Comparing temporal auto-correlations of velocity, when flows are nonrotating, between magnetic (solid line) and 
non-magnetic (dashed line) flows.
}
\label{tempcomnorot}
\end{figure}


\begin{figure}[H]
 \centering
\includegraphics[scale=0.9]{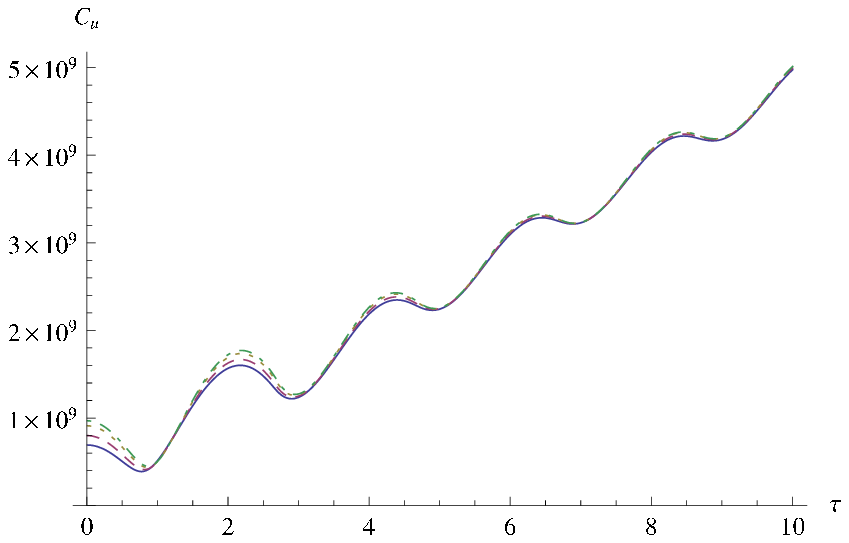}
\caption{Temporal auto-correlations of velocity, when $q=1.5$ (solid line), $1.7$ (dashed line), $1.9$ (dotted line), $1.9999$ (dotdashed line).
}
\label{tempallfiniteq}
\end{figure}

First of all, we show in Figures \ref{tempcomq1p5} and \ref{tempcomnorot} that independent of the effects due to 
rotation and noise, the presence of magnetic field solely increases the auto-correlation enormously compared
to that in the absence of magnetic field. This establishes the power of magnetic field and associated
Alfv\'en wave in modulating the growth of perturbation. This further establishes the importance of the 
present work over that by Mukhopadhyay \& Chattopadhyay \cite{mc13}.

Now we concentrate on magnetized flows.
Figure \ref{tempallfiniteq} shows that in the magnetic flows with the decrease of $q$, although the 
velocity correlation decreases, the difference between those of
any two $q$s is insignificant. Hence, the simultaneous presence of magnetic field and stochastic white noise kills the 
dependence of correlations on the rotational effect.
\begin{figure}[H]
 \centering
\includegraphics[scale=0.9]{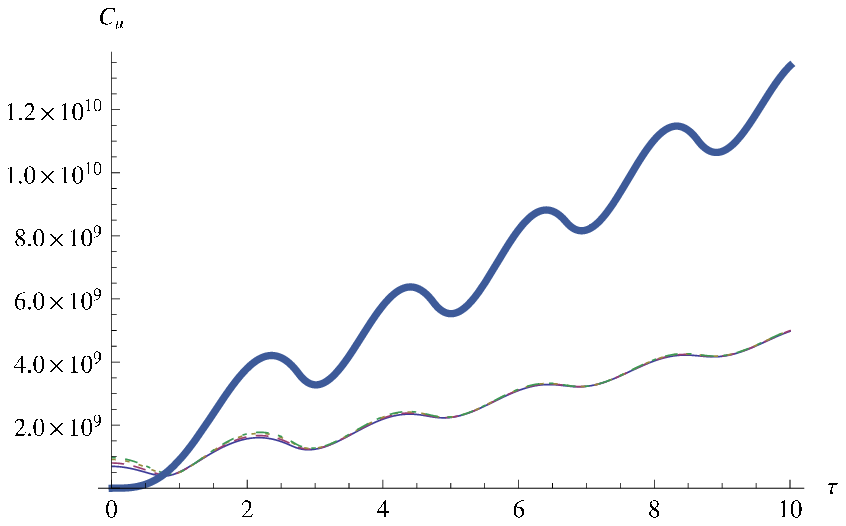}
\caption{Same as Figure 1, but auto-correlation for nonrotating flow (thick-solid line) is additionally shown.}
\label{tempallq}
\end{figure}
\begin{figure}[H]
 \centering
\includegraphics[scale=0.9]{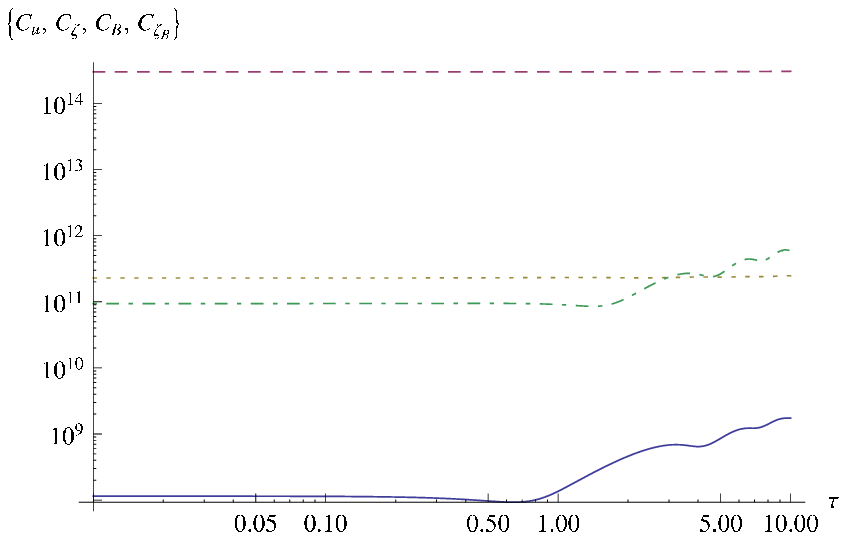}
\caption{Temporal auto-correlations of velocity (solid line), vorticity (dashed line), magnetic field (dotted line) and 
magnetic vorticity (dot-dashed line), when $q=1.5$.}
\label{tempautoallq15}
\end{figure}
Note that in the absence of noise, hydrodynamic perturbation energy growth (and hence
$C_u(\tau)$) in all the above rotating cases, essentially for $q<2$, is very small, as shown previously \cite{man}, 
particularly in three dimensions. However, the effects of magnetic field and noise bring
in a huge growth of the perturbation top of the Coriolis fluctuations, clearly revealing instability. 
A remarkable feature in the scaling nature of all these correlations 
is their independence of $q$ (background angular velocity profile) --- 
a trait identified in statistical physics literature as {\it universality}.

In Figure \ref{tempallq}, we show that the correlation for a nonrotating magnetized flow 
appears to be quite larger compared to that for rotating flows, which is similar to the
trait observed in the absence of noise (see, e.g., \cite{bmraha,man}). However, the presence of noise increases
the growth of perturbations enormously in either of the cases. 

Figure \ref{tempautoallq15} depicts that the auto-correlation for vorticity perturbation is
largest among all the auto-correlations for a particular $q$. Note also that 
auto-correlations for velocity and magnetic vorticity exhibit more oscillations 
compared to that for magnetic field and vorticity. This is because the effects due
to Alfv\'en wave arised from magnetic field. Note from equations (\ref{orrv}) and 
(\ref{orrbzeta}) that the evolutions of velocity and magnetic vorticity depend on 
the magnetic perturbation explicitly, and hence the respective auto-correlations get modulated
by Alfv\'en waves. Moreover, the amplitude of velocity correlation is smallest
at the beginning due to fluctuations arised in the velocity perturbation, 
whose curl however need not be small, giving rise to large vorticity correlations.
However, either of correlations is large enough to govern instability and then turbulence. 
Nevertheless, all the correlations saturate (or tend to saturate) at a relatively
large $\tau$ (which is more clearer in the log-linear plots described in \S V).

\subsubsection{Cross-Correlations}

Here we stick to the same assumptions as of the computations of auto-correlations.
The temporal cross-correlation of two quantities, e.g. $u$ and $\zeta$, is defined by
\begin{eqnarray}
<u(\vec{x},t)\,\zeta(\vec{x},t+\tau)>=C_{u\zeta}(\tau)=\int d^3k\,d\omega\,e^{-i\omega\tau}<\tilde{u}_{\vec{k},\omega}\,
\tilde{\zeta}_{-\vec{k},-\omega}>.
\label{tempcrosscorr}
\end{eqnarray}
Similarly, one can define other cross-correlations. We solve the integrals following the same procedure
as described for auto-correlations.

\begin{figure}[H]
 \centering
\includegraphics[scale=0.9]{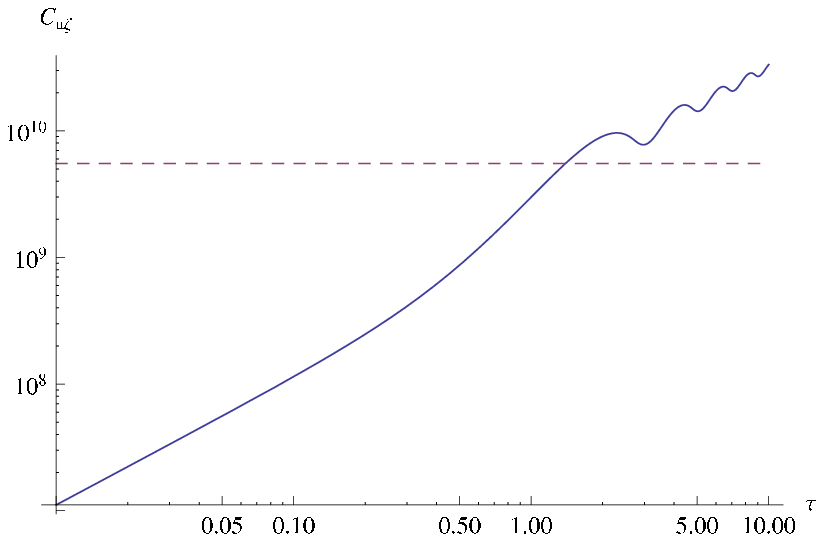}
\caption{Comparing temporal cross-correlations of velocity and vorticity, 
when $q=1.5$, between magnetic (solid line) and non-magnetic (dashed line) flows.
}
\label{temp_cross_comp}
\end{figure}

First of all, we show in Figure \ref{temp_cross_comp} that unlike auto-correlations, the cross-correlation
of velocity and vorticity decreases quite a bit in the presence of magnetic field compared to that in the absence of it
at the beginning.
This further pinpoints the additional effects arised due to the magnetic field.

\begin{figure}[H]
 \centering
\includegraphics[scale=0.9]{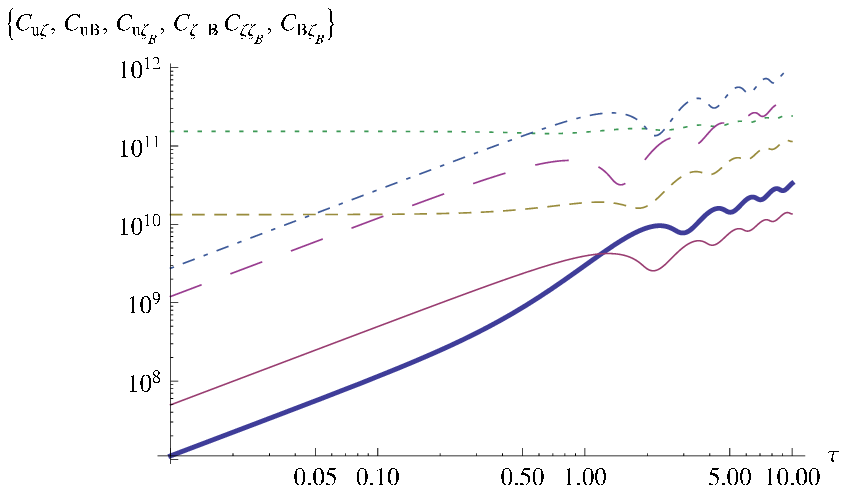}
\caption{Temporal cross-correlations of velocity and vorticity (thick solid line), velocity and magnetic field (solid line), 
velocity and magnetic vorticity (dashed line), vorticity and magnetic field (dotted line), vorticity and magnetic vorticity (dot-dashed line), magnetic field and magnetic vorticity (long dashed line), when $q=1.5$.}
\label{temp_cross_q15}
\end{figure}

Figure \ref{temp_cross_q15} shows all the cross-correlations in a magnetized Keplerian disk. 
Interestingly, cross-correlations of velocity and magnetic vorticity (dashed line) and
vorticity and magnetic field (dotted line) have a steady, constant, higher amplitude at the 
beginning compared to other cross-correlations. This is because they are correlations
of either two fluctuating (due to Alfv\'en wave) variables or two non-fluctuating variables,
when, as shown in Figure \ref{tempautoallq15} that, one of them have larger amplitude
to begin with. 
All the remaining ones are the correlations of a strongly Alfv\'en wave modulated variable
with a non-modulated variable.
Because of the same reason, the velocity-vorticity cross-correlation in the non-magnetized flow
is larger than that in the magnetized flow, when magnetic field modulates velocity perturbation
but not the vorticity perturbation.

\subsection{Spatial correlation}
\subsubsection{Auto-correlations}

Here also we assume
$<\tilde{{\eta}_i}_{\vec{k},\omega}\,\tilde{{\eta}_j}_{-\vec{k},-\omega}>=\delta_{ij}$, like the case of temporal correlations, and 
obtain spatial correlations of
velocity, vorticity, magnetic field and magnetic vorticity, given below as
\begin{eqnarray}
\nonumber
&&<u(\vec{x},t)\,u(\vec{x}+\vec{r},t)>=S_u(r)=\int d^3k\,d\omega\,e^{i\vec k.\vec r}
<\tilde{u}_{\vec{k},\omega}\,
\tilde{u}_{-\vec{k},-\omega}>,\\
\nonumber
&&<\zeta(\vec{x},t)\,\zeta(\vec{x}+\vec{r},t)>=S_\zeta(r)=\int d^3k\,d\omega\,e^{i\vec k.\vec r}
<\tilde{\zeta}_{\vec{k},\omega}\,
\tilde{\zeta}_{-\vec{k},-\omega}>,\\
\nonumber
&&<B_x(\vec{x},t)\,B_x(\vec{x}+\vec{r},t)>=S_{B_x}(r)=\int d^3k\,d\omega\,e^{i\vec k.\vec r}
<\tilde{B_x}_{\vec{k},\omega}\,
\tilde{B_x}_{-\vec{k},-\omega}>,\\
\nonumber
&&<\zeta_B(\vec{x},t)\,\zeta_B(\vec{x}+\vec{r},t)>=S_{\zeta_B}(r)=\int d^3k\,d\omega\,e^{i\vec k.\vec r}
<\tilde{\zeta_B}_{\vec{k},\omega}\,
\tilde{\zeta_B}_{-\vec{k},-\omega}>.\\
\label{velzetmagcorr}
\end{eqnarray}

Now using equations (\ref{mat1}) and (\ref{velzetmagcorr}), the 
spatial correlation of velocity perturbation $S_u(r)$ 
is explicitly given by

\begin{eqnarray}
&&S_u(r) 
= 2\pi\int_{k_0}^{k_m}~dk~k^2 ~\int_0^\pi~d\theta~e^{ikr\cos\theta}~
\int d\omega~<\tilde{u}_{\vec{k},\omega}\,
\tilde{u}_{-\vec{k},-\omega}>,
\label{spatialcorr}
\end{eqnarray}

\noindent
where the $\theta-$integral is the zeroth-order Bessel function $J_0(kr)$. Similarly, one can
obtain $S_\zeta(r)$, $S_{B_x}(r)$, $S_{\zeta_B}(r)$ explicitly, when the poles of the 
integrand of equation 
(\ref{spatialcorr}) and of equations for other correlations are identified. 
Here also we stick to the simplifying assumption $k_x=k_y=k_z=k/\sqrt{3}$. 


\begin{figure}[H]
 \centering
\includegraphics[scale=0.9]{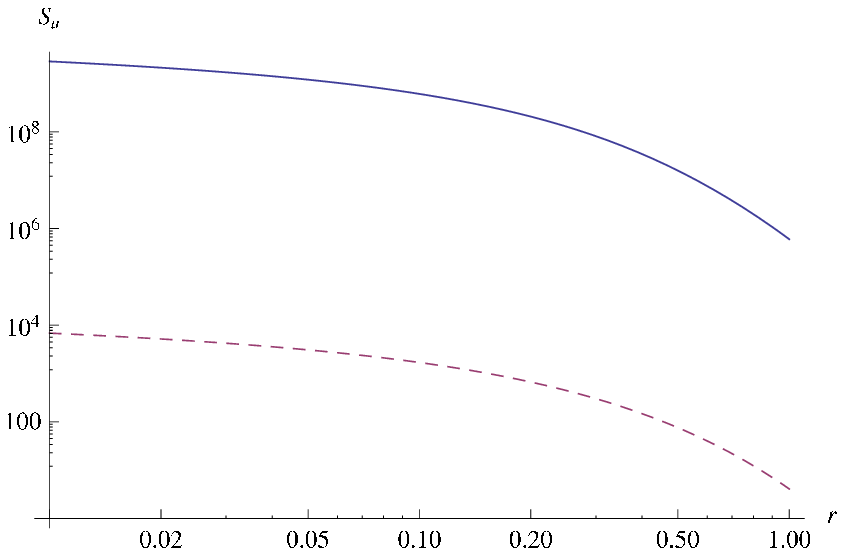}
\caption{Comparing spatial auto-correlations of velocity, when $q=1.5$, between magnetic (solid line) and 
non-magnetic (dashed line) flows.
}
\label{spatcomq1p5}
\end{figure}

\begin{figure}[H]
 \centering
\includegraphics[scale=0.9]{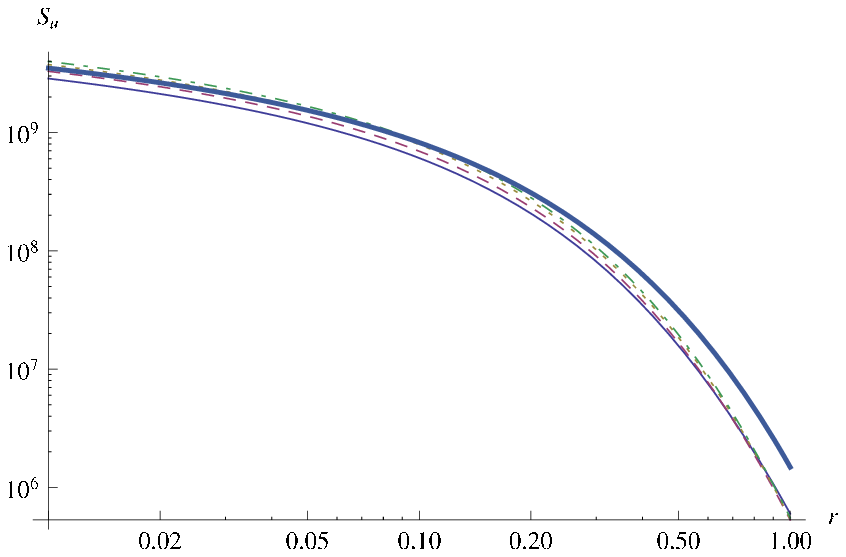}
\caption{Spatial auto-correlations of velocity, when $q=1.5$ (solid line), $1.7$ (dashed line), $1.9$ (dotted line), $1.9999$ (dotdashed line) and flow is nonrotating (thick-solid line).}
\label{spatialallq}
\end{figure}

Like the discussions of temporal correlations, here also we begin by comparing the results between 
magnetic and non-magnetic flows, as shown in Figure \ref{spatcomq1p5}. This again confirms that the 
magnetic field creates an additional effect leading to a much larger growth of perturbation.

Figure \ref{spatialallq} shows that the spatial correlations of velocity perturbation in magnetized flows
decrease with the decrease of $q$ from $1.9999$, while the difference between those of any two $q$s
is insignificant. Note that the nonrotating case gives a slightly larger correlation
than all the rotating cases.
It is generally seen that the correlations decrease with increasing $r$ as well.
However, their value appears significant enough
to reveal a steadily damped instability in the flow.
Such large values of perturbation energy growth are indicative of 
instability and plausible turbulent transport, in the presence of stochastic noise.

\begin{figure}[H]
 \centering
\includegraphics[scale=0.9]{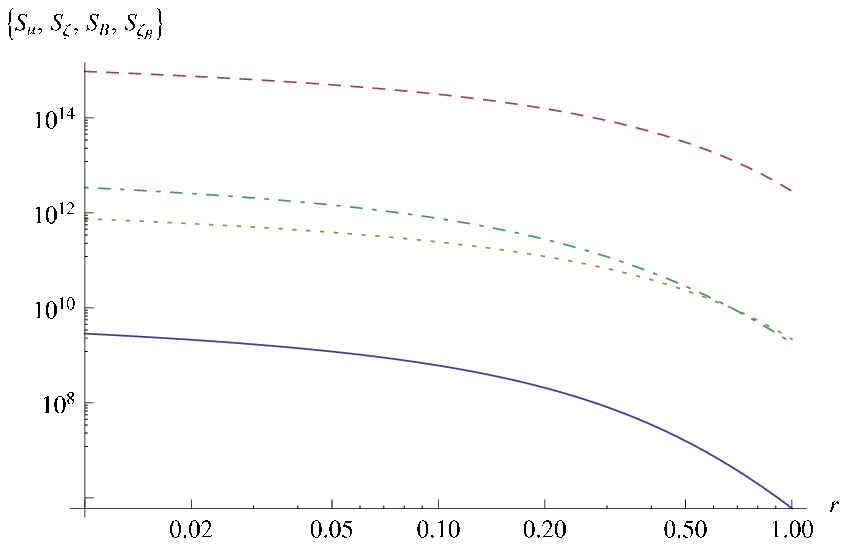}
\caption{Spatial auto-correlations of velocity (solid line), vorticity (dashed line), magnetic field (dotted line), magnetic vorticity (dot-dashed line), when $q=1.5$.}
\label{Spatial_autocorrelations_q15}
\end{figure}

Figure \ref{Spatial_autocorrelations_q15} shows all spatial auto-correlations 
for a magnetized Keplerian disk. Like the temporal case, velocity correlation is lowest
and its curl, i.e. the vorticity correlation, is highest. The underlying reasons 
being similar as described in the case of temporal correlations, when the existence of 
modulation due to Alfv\'en wave plays a determining rule. However, all the correlations
are large enough to reveal instability.

\subsubsection{Cross-Correlations}

Here we stick to the same assumptions as of the computations of auto-correlations.
We can define the spatial cross-correlation of two quantities, e.g. $u$ and $\zeta$, as
\begin{eqnarray}
&&<u(\vec{x},t)\,\zeta(\vec{x}+\vec{r},t)>=S_{u\zeta}(r)=\int d^3k\,d\omega\,e^{i\vec k.\vec r}
<\tilde{u}_{\vec{k},\omega}\,
\tilde{\zeta}_{-\vec{k},-\omega}>.
\label{spatialcrosscorr}
\end{eqnarray}
Similarly, one can define other cross-correlations.

\begin{figure}[H]
 \centering
\includegraphics[scale=0.9]{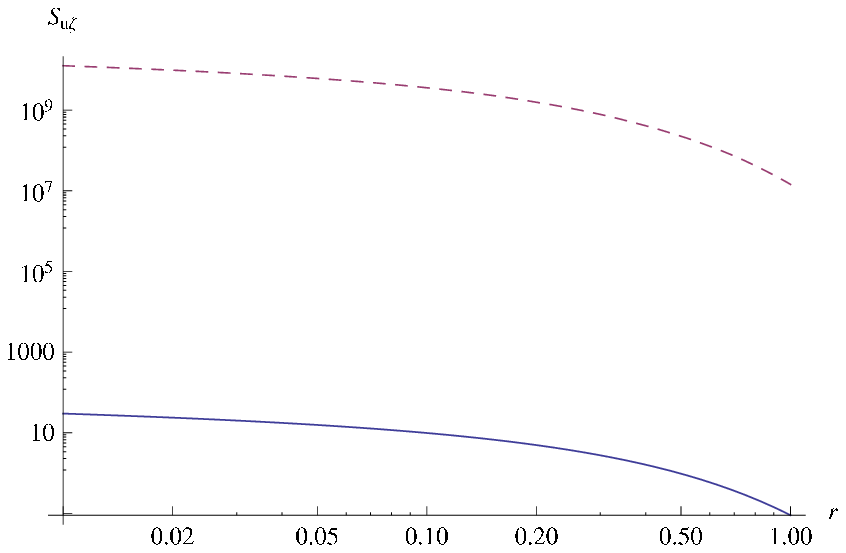}
\caption{Comparing spatial cross-correlations of velocity and vorticity, 
when $q=1.5$, between magnetic (solid line) and non-magnetic (dashed line) flows.
}
\label{spat_cross_comp}
\end{figure}

Figure \ref{spat_cross_comp} further confirms that the cross-correlations can 
behave in the opposite fashion in the presence of magnetic field, compared to the auto-correlation.
The reason being the same as that discussed in order to describe the temporal cross-correlations.

\begin{figure}[H]
 \centering
\includegraphics[scale=0.9]{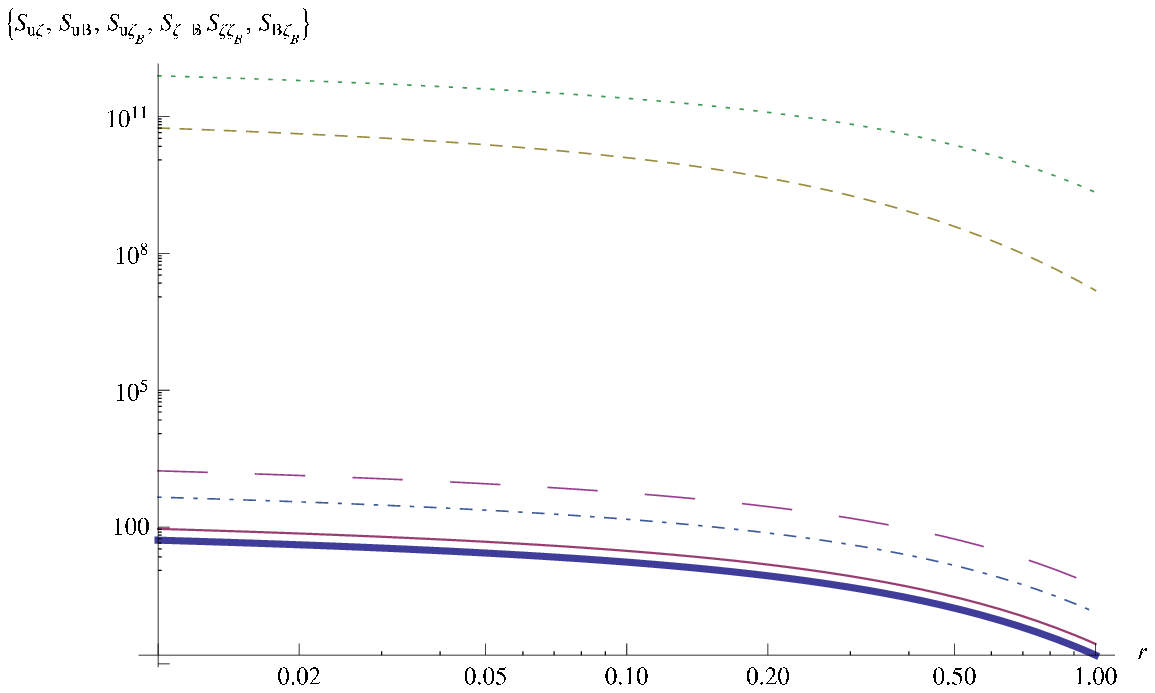}
\caption{Spatial cross-correlations of velocity and vorticity (thick solid line), velocity and magnetic field (solid line), velocity and magnetic vorticity (dashed line), vorticity and magnetic field (dotted line), vorticity and 
magnetic vorticity (dot-dashed line), magnetic field and magnetic vorticity (long dashed line), when $q=1.5$.}
\label{Spatial_cross_q15}
\end{figure}

Figure \ref{Spatial_cross_q15} shows all spatial cross-correlations for a magnetized Keplerian disk,
which is in accordance with auto-correlations described in Figure \ref{Spatial_autocorrelations_q15}
and the description for temporal cross-correlations.

\section{Two-point correlations of perturbation in the presence of colored noise}

\begin{figure}[H]
 \centering
\includegraphics[scale=0.9]{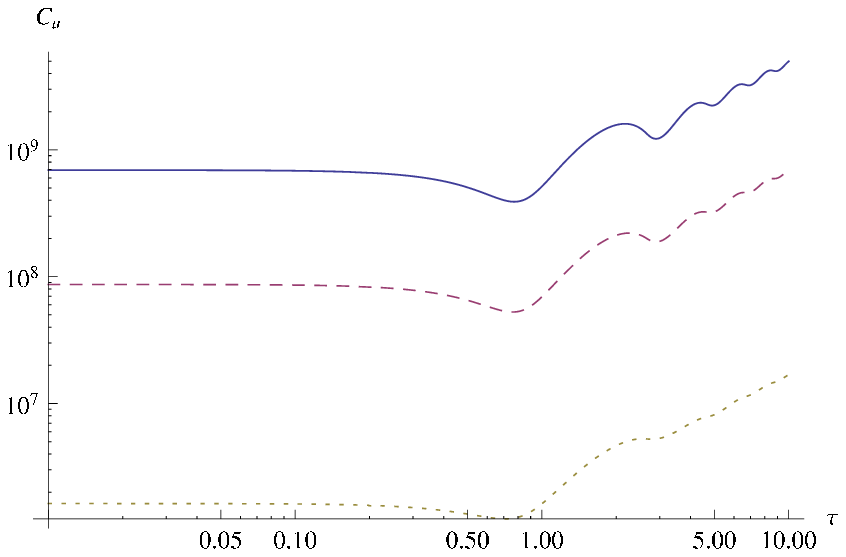}
\caption{Temporal auto-correlations of velocity for $q=1.5$, when $D_i=k^0$ (solid line), $k^{-1}$ 
(dashed line), $k^{-3}$ (dotted line).
}
\label{temvelcol}
\end{figure}

\begin{figure}[H]
 \centering
\includegraphics[scale=0.9]{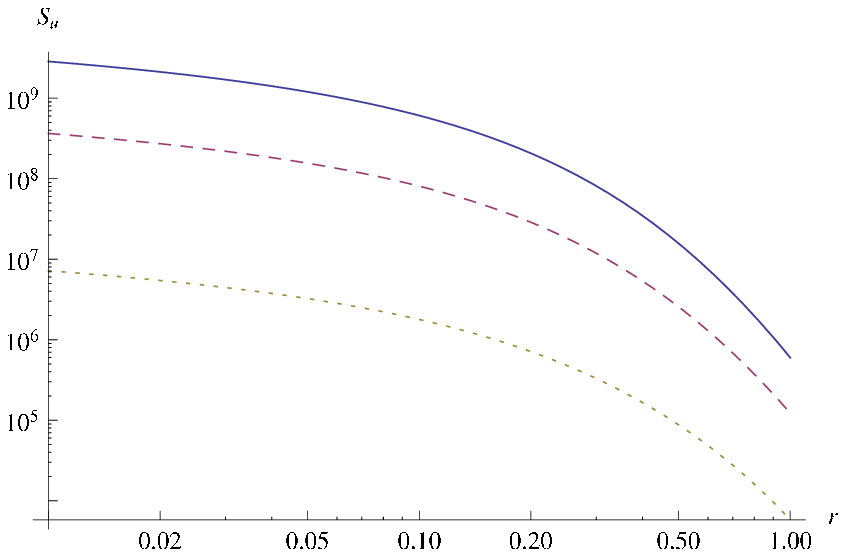}
\caption{Spatial auto-correlations of velocity for $q=1.5$, when $D_i=k^0$ (solid line), $k^{-1}$ 
(dashed line), $k^{-3}$ (dotted line).
}
\label{spvelcol}
\end{figure}

\begin{figure}[H]
 \centering
\includegraphics[scale=0.9]{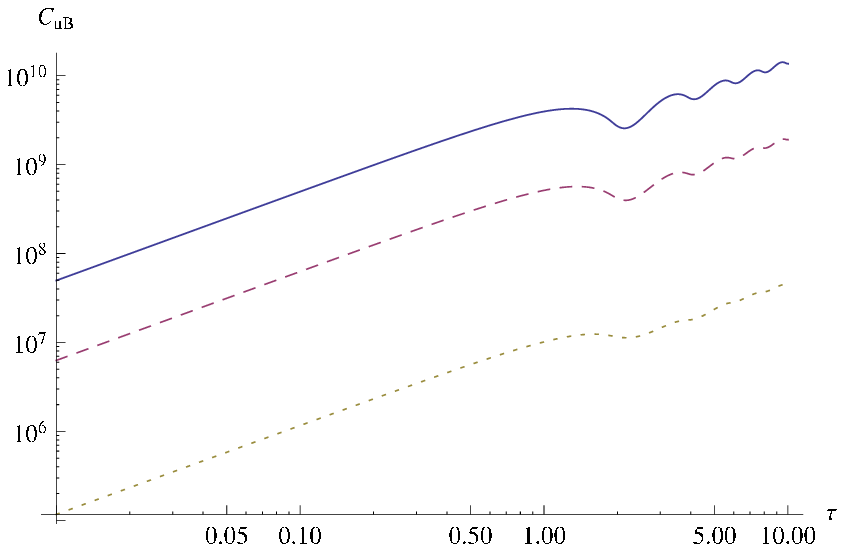}
\caption{Temporal cross-correlations of velocity and magnetic field 
for $q=1.5$, when $D_i=k^0$ (solid line), $k^{-1}$ 
(dashed line), $k^{-3}$ (dotted line).
}
\label{temvelbcol}
\end{figure}

\begin{figure}[H]
 \centering
\includegraphics[scale=0.9]{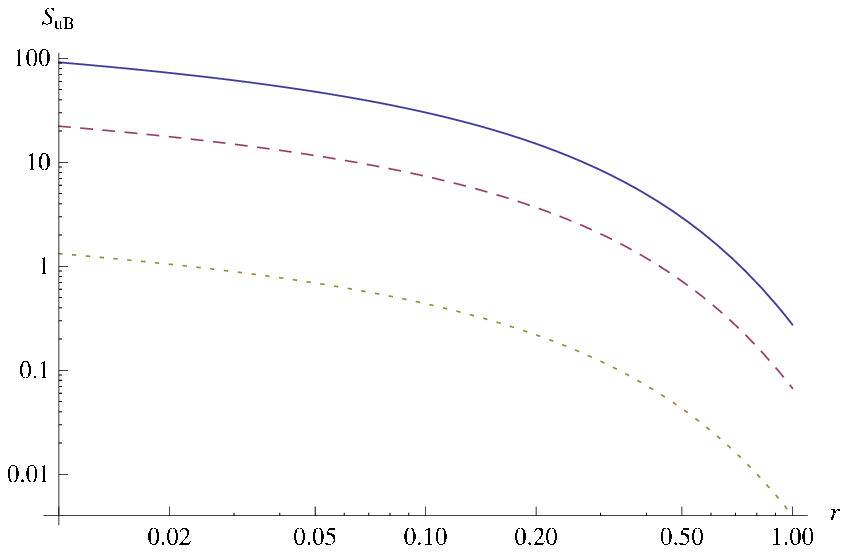}
\caption{Spatial cross-correlations of velocity and magnetic field 
for $q=1.5$, when $D_i=k^0$ (solid line), $k^{-1}$ 
(dashed line), $k^{-3}$ (dotted line).
}
\label{spvelbcol}
\end{figure}

Here we show, how the effects of colored noise change the correlations, mainly
their amplitudes. For this purpose, we stick to a particular background profile
which corresponds to the Keplerian disk. We consider the colored noise in such
a way that the correlation function $D_i$ scales as $1/k^{3-\alpha}$. Then 
we choose three values of $\alpha$, which are $3$ (white noise), $2$ and $0$
(no vertex correction). In Figures \ref{temvelcol} and \ref{spvelcol}, 
we compare effects of various colored noise to the temporal and spatial auto-correlations 
respectively. Further, in Figures \ref{temvelbcol} and \ref{spvelbcol}, we compare 
effects of various colored noise to a typical (velocity and magnetic field)
temporal and spatial cross-correlations respectively. The figures clearly show
that effects of colored noise decrease the correlations --- larger the magnitudes of 
slop of $D_i$, smaller the correlations are. However, even for $D_i\sim k^{-3}$,
auto-correlations are large enough to govern instability.

\section{Summary and conclusions}

In this work, we have attempted to address the origin of instability and then turbulence in 
magnetized, rotating, shear flows (more precisely a small section of it, which is a plane shear 
flow supplemented by the Coriolis force). Our particular emphasis is
the flows having decreasing angular velocity but increasing specific angular momentum with the radial
coordinate, which are Rayleigh stable. The flows with such a kind of velocity profile are 
often seen in astrophysics. 
As the molecular viscosity in astrophysical accretion disks
is negligible, any transport of matter therein would arise through turbulence only, in order
to explain observed data.
Therefore, essentially we have addressed here the
plausible origin of viscosity in rotating shear flows of the kind mentioned above. 
Note that whether a flow is magnetically arrested or not,
hydrodynamic effects always exist, as Mukhopadhyay \& Chattopadhyay \cite{mc13} argued. 
Hence, the relative strengths of hydrodynamics and hydromagnetics 
in the time scale of interest determines the actual source of instability. Present work shows that
the strength of hydromagnetic effects could be superior than that of hydrodynamic effects.

We have shown, based on the theory of statistical physics (which has been recalled in 
detail by Mukhopadhyay \& Chattopadhyay \cite{mc13}, in the present context), that 
stochastically forced linearized rotating shear flows 
in a narrow gap limit reveal a very large correlation energy growth of
perturbation in the presence of magnetic field and noise.
We have shown separately (1) the sole effects of magnetic field
at fixed noise and rotation, (2) sole effects of rotation at fixed magnetic
field and noise, and (3) sole effects of noise at fixed magnetic field and
rotation. 

Although the correlations of perturbation decrease as the flow deviates
from the type with $q=1.9999$ (when $q=2$ exactly corresponds to constant specific angular momentum) 
to that of the Keplerian, the difference 
between them is very small
and they appear large enough to trigger nonlinear effects and instability. 

\begin{figure}[H]
 \centering
\includegraphics[scale=0.9]{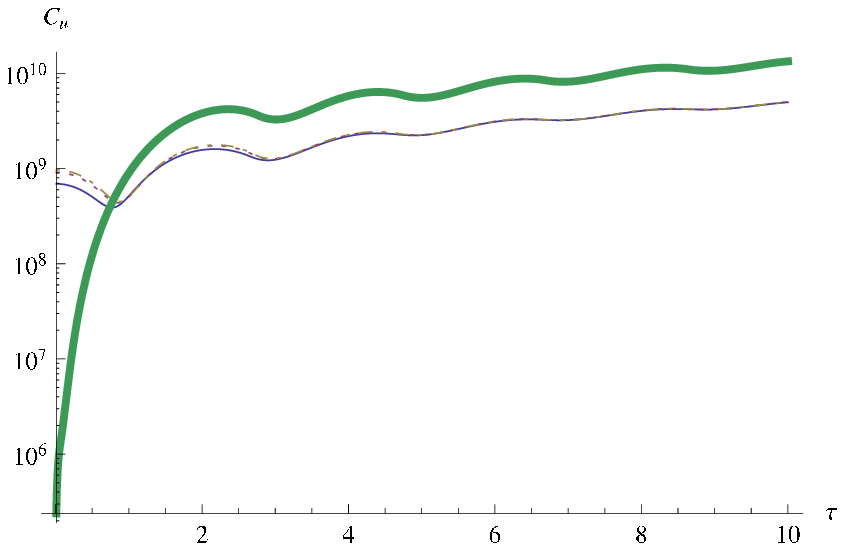}
\caption{Temporal auto-correlations, in log-linear scale, of velocity, when $q=1.5$ (solid line),  
$1.9$ (dashed line), $1.9999$ (dotted line) and flow is nonrotating (thick-solid line).
}
\label{loglinauto}
\end{figure}

\begin{figure}[H]
 \centering
\includegraphics[scale=0.9]{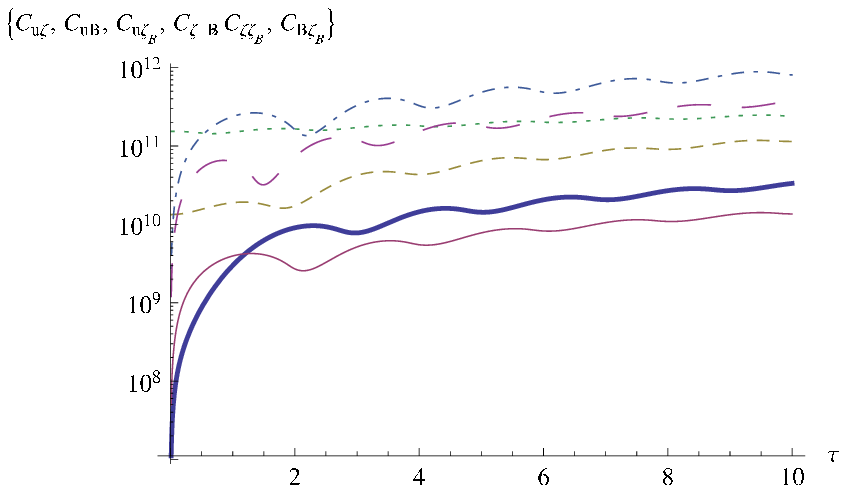}
\caption{Temporal cross-correlations, in log-linear scale, of velocity and vorticity (thick solid line), velocity and magnetic field (solid line), 
velocity and magnetic vorticity (dashed line), vorticity and magnetic field (dotted line), vorticity and magnetic vorticity (dot-dashed line), magnetic field and magnetic vorticity (long dashed line), when $q=1.5$.}
\label{loglincros}
\end{figure}

Therefore, the present work addresses the 
{\it large three-dimensional hydromagnetic energy growth of linear perturbation}, 
in the line with theoretical framework grounded by Mukhopadhyay \& Chattopadhyay \cite{mc13},
which presumably leads to instability and subsequent turbulence.
Only requirement here
is the presence of stochastic noise and magnetic field in the system together, which is quite 
obvious in natural flows 
like astrophysical (hot) accretion disks around compact objects.
Interestingly, all the flows with $q<2$, exhibiting 
very similar
growth and roughness exponents with almost identical energy dissipation amplitudes, indicates the 
universality class. In addition, all the correlations tend to saturate at a large time. 
This feature is clearer in the log-linear plots given by Figures \ref{loglinauto} and \ref{loglincros}.
It is known that the time for maximum transient growths arised in a non-normal system, like the one
under consideration, scales as $R_e^\gamma$, when $\gamma\ge0$ \cite{man}. As $R_e$ and $R_m$ are chosen to be very large in the 
present work, in practical time scales the correlations never reveal any transient behavior. Note that, 
at large $r$, all the spatial correlations reveal similar amplitude as well.
Thus the properties of temporal and spatial correlations together, in the presence of noise and 
magnetic field, indicate that
the Rayleigh stable rotating shear flows follow a single universality class. 
Another aspect to be noted from our work is that the presence of magnetic field brings in oscillatory 
nature in the energy growths of the system (due to the presence of Alfv\'en wave), unlike the 
energy growths in hydrodynamic case \cite{mc13}. Therefore, there might be a possibility of 
existence of a flow in which the presence of magnetic field hinders the energy growth of perturbation instead 
of enhancing the same --- a veritable destructive interference. 
This, however, has to be investigated in detail, in particular relaxing the 
choice of specific wave-vector of perturbations and also including dominant non-linear perturbing modes.

\section*{Acknowledgments}
The authors thank Bruno Eckhardt for insightful comments and suggestions over the primary 
version of the paper. Thanks are also due to the referees for their comments to
improve the presentation of the paper.
B.M. thanks Indian Space Research Organization (ISRO) project, grant number ISRO/RES/2/367/10-11, 
for a partial support.
A.K.C. thanks the Royal Society, U.K., research grant number RG110622, for partial 
support.

{}

\end{document}